\documentclass[epj]{svjour}
\usepackage{graphics}
\usepackage{hyperref}
\hypersetup{colorlinks = true, allcolors = blue}
\usepackage[utf8]{inputenc}
\usepackage{authblk}
\usepackage{hepparticles}
\usepackage{hepunits}
\usepackage{hepnames}
\usepackage{lineno}
\usepackage{multirow}
\begin{document}
\title{Calorimetry at FCC-ee}
%\subtitle{If you have a sub-title, enter it here}
\author{Martin Aleksa\inst{1} 
\and Franco Bedeschi\inst{2} 
\and Roberto Ferrari\inst{3} 
\and Felix Sefkow\inst{4} 
\and Christopher G.\ Tully\inst{5}% etc
}                     % Do not remove
\offprints{}          % Insert a name or remove this line
\institute{CERN, EP Department, Geneva, Switzerland \and INFN, Sezione di Pisa, Italy \and INFN, Sezione di Pavia, Italy \and  Deutsches Elektronen-Synchrotron (DESY), Hamburg, Germany \and Princeton University, Department of Physics, Princeton, NJ, United States}
\date{Received: \today / Revised version: \today }
\abstract{
%Insert your abstract here.
With centre-of-mass energies covering the Z pole, the WW threshold, the HZ production, and the top-pair threshold, the FCC-ee offers unprecedented possibilities to measure the properties of the four heaviest particles of the Standard Model (the Higgs, Z, and W bosons, and the top quark), and also those of the b and c quarks and of the $\tau$ lepton. At these moderate energies, the role of the calorimeters is to complement the tracking systems in an optimal (a.k.a. particle-flow) event reconstruction. In this context, precision measurements and searches for new particles can fully profit from the improved electromagnetic and hadronic object reconstruction offered by new technologies, finer transverse and longitudinal segmentation, timing capabilities, multi-signal readout, modern computing techniques and algorithms. The corresponding requirements arise in particular from the resolution on reconstructed hadronic masses, energies, and momenta, e.g., of H, W, Z, needed to reach the FCC-ee promised precision. Extreme electromagnetic energy resolutions are also instrumental for $\pi^0$ identification, $\tau$ exclusive decay reconstruction, and physics sensitivity to processes accessible via radiative return. We present state of the art, challenges and future developments on some of the currently most promising technologies: high-granularity silicon and scintillator readout, dual readout, noble-liquid and crystal calorimeters.
\PACS{
      {PACS-key}{describing text of that key}   \and
      {PACS-key}{describing text of that key}
     } % end of PACS codes
} %end of abstract
\maketitle

\section{Introduction}
%\label{sec:intro}
%The Future Circular Collider (FCC) is an ambitious project of an accelerator complex in the CERN area for the era after LHC. Whereas an electron-positron collider, FCC-ee~\cite{Benedikt:2651299}, is considered as a possible first step to measure precisely the Higgs properties, the tunnel length and infrastructure of the complex are defined by a 100\,TeV hadron circular collider, the FCC-hh~\cite{Benedikt:2651300}. 

\label{sec:intro}
The Future Circular Collider (FCC) is an ambitious project of an accelerator complex in the CERN area for the era after LHC~\cite{Benedikt:2653673}.  An electron-positron collider, FCC-ee~\cite{Benedikt:2651299}, is considered as a possible first step to precisely measure the Higgs properties, improve by orders of magnitude the measurement of key electroweak parameters and complement the study of heavy flavours of Belle2~\cite{Kou:2018nap} and LHCb\cite{Alves:2008zz,Bediaga:2018lhg}. 

This vast physics program relies in many ways on the calorimeters, whose performance is enhanced by the inclusion of the tracking information with particle-flow methods.  In particular, calorimeters must provide precise hadronic jet measurements in two- or more-jet final states.  Z, W or Higgs decays into two jets have the largest branching fractions for each of the bosons, and ZZ and WW final states represent the major backgrounds to most Higgs-boson decay modes.  A $\sim$3-4\% two-jet invariant-mass resolution is needed to adequately classify all relevant final states. This is a hard requirement on the hadronic calorimetry that cannot be achieved with conventional methods.  New detector concepts that can meet these goals are high-granularity tracking calorimeters, optimised for particle flow (PFlow), and dual-readout calorimeters.  Recently, it has been shown that calorimeters based on cryogenic liquids with high readout granularity can also be optimised for particle-flow and 4D imaging techniques and hence become viable choices.  

%New reconstruction and analysis tools based on machine learning as well as a large increase in read-out granularity have been shown to improve hadronic calorimeter resolution in many contexts, thus making also more conventional technologies, like those based on cryogenic liquids, viable choices.

The requirements on electromagnetic (EM) calorimeters are mainly driven by the need of a good $\pi^0$ reconstruction, that is relevant for the identification of specific $\tau$-lepton or heavy-flavoured-hadron final states. Physics sensitivity to some processes accessible via radiative return also requires a very good EM resolution.  Best performances are given by technologies based on cryogenic noble liquids or crystals, the latter providing extreme EM resolution.

In the following we present the current status and prospects for all the calorimeter technologies relevant for FCC-ee and discuss their key R\&D issues.

\section{Highly Granular Silicon and SiPM-Scintillator-Tile Calorimeters }
\label{sec:HG-Si-Scint}
The idea to apply the particle-flow approach in a future $e^+e^-$ collider detector, for the precision study of heavy particles predominantly decaying into jets, has driven the development of highly granular calorimeters from the beginning~\cite{Brient:2002gh,Morgunov:2001cd}. 
The PFlow method optimises the jet energy resolution by individually reconstructing each particle and using the best measurement for each, technique which poses high demands on the imaging capabilities of calorimeters.
Charged particles are best measured with tracking detectors and photon energies can be measured with a relative precision of about $15\%/\sqrt{E(\rm{GeV})}$, or better, in electromagnetic calorimeters. 
In a typical jet, about 60\% of the energy is carried by charged particles, 30\% by photons and only 10\% by long-lived neutral hadrons ($K^0_L$ and neutrons), for which detection hadronic calorimetry is imperative. 
Assuming a hadronic energy resolution of $55\%/\sqrt{E(\rm{GeV})}$, then, if ideally each particle is resolved, a jet energy resolution of $19\%/\sqrt{E(\rm{GeV})}$ could be obtained where the dominant part ($17\%/\sqrt{E(\rm{GeV})}$) is still due to the calorimeter resolution for neutral hadrons.

In practice, mis-assignments give rise to additional measurement uncertainties, called {\it confusion}\index{confusion} term. 
For example, a neutral particle shower could be misinterpreted as part of a nearby charged hadron shower, and the neutral energy would be lost, or a detached fragment of a charged particle shower could be misidentified as a separate neutral hadron, and the fragment energy would be double counted. 

Particle-flow calorimeters, with their emphasis on imaging, must still feature a good hadron energy resolution.  The neutral-hadron energy uncertainty is the dominant contribution to the jet resolution for low energy jets, where particles are well separated. At higher energies, the confusion effects take over, and a good calorimetric resolution improves the energy-momentum match used in the assignment of energy depositions. For the typical particle-flow driven $e^+e^-$ detectors, the transition is at jet energies around 100~GeV.

The principle has been experimentally tested~\cite{Sefkow:2015hna} by the CALICE collaboration with test beam prototypes using different absorber materials and readout techniques. 
The most commonly proposed technologies are silicon diodes for the electromagnetic section and scintillating tiles, individually read out by silicon photomultipliers (SiPMs), for the hadronic part. 
Scintillator ECAL and gaseous HCAL technologies are being explored, too.

In all cases, the high channel density requires the integration of the front-end electronics into the active layers, such that the digitised and zero-suppressed data can be extracted from the volume via a small number of readout lines.
The initial focus on applications at linear colliders, with their low duty cycle, has helped to keep the associated requirements for data transfer, power and cooling, manageable. 
These need to be revised for an implementation at FCC-ee, with possible implications on the overall calorimeter architectures and integration concepts.

\subsection{State of the Art}
A first systematic optimisation of the 3D detector segmentation parameters, using detailed simulations and the PANDORA reconstruction algorithm, has been done in~\cite{Thomson:2009rp}, for jet energies up to 250~GeV, and later confirmed~\cite{Tran:2017tgr}, for the HCAL with software compensation taken into account.  The proposed HCAL cell sizes are of the order of a radiation length, the characteristic scale of shower sub-structure, about 3~cm in a steel-plastic structure. In concepts with 1- or 2-bit ((semi-) digital) readout, 1~cm cells are needed. 
ECAL cell sizes of typically 0.5~cm, well below 1~$X_0$, were shown to provide superior separation power for nearby electromagnetic showers in their early evolution stage, before they attain their full width. 

The first generation prototypes built by the CALICE collaboration did not yet have fully integrated front-end electronics, but were successfully used to validate the  particle separation power~\cite{Adloff:2011ha} and the single-particle energy resolutions expected from simulations. 
For the ECAL, 
a stochastic term of 
$16.5\%/\sqrt{E({\rm GeV})}$ was measured in the range 6 - 45~GeV with a constant term of 1.1\%~\cite{Anduze:2008hq}.
%\begin{equation*}
%\label{eq:datares}
% \sigma (E_{\mathrm{meas}})/E_{\mathrm{meas}}   = 
% (16.53\pm0.14(\mathrm{stat})\pm0.4%(\mathrm{syst}) )\%/\sqrt{E(\mathrm{GeV})}  
%\oplus   \left(1.07\pm0.07(\mathrm{stat})\pm0.1(\mathrm{syst})\right)\%
%\end{equation*}
For the HCAL, $44.2\%/\sqrt{E({\rm GeV})}$ and 1.8\% were measured in the range 10 - 80~GeV~\cite{Adloff:2012gv}, using a cell-energy based weighting procedure.

\begin{figure}
    \centering
%    \sidecaption
%\includegraphics[width=0.5\textwidth]{./slab_test.jpg}  
\resizebox{0.54\textwidth}{!}{\includegraphics{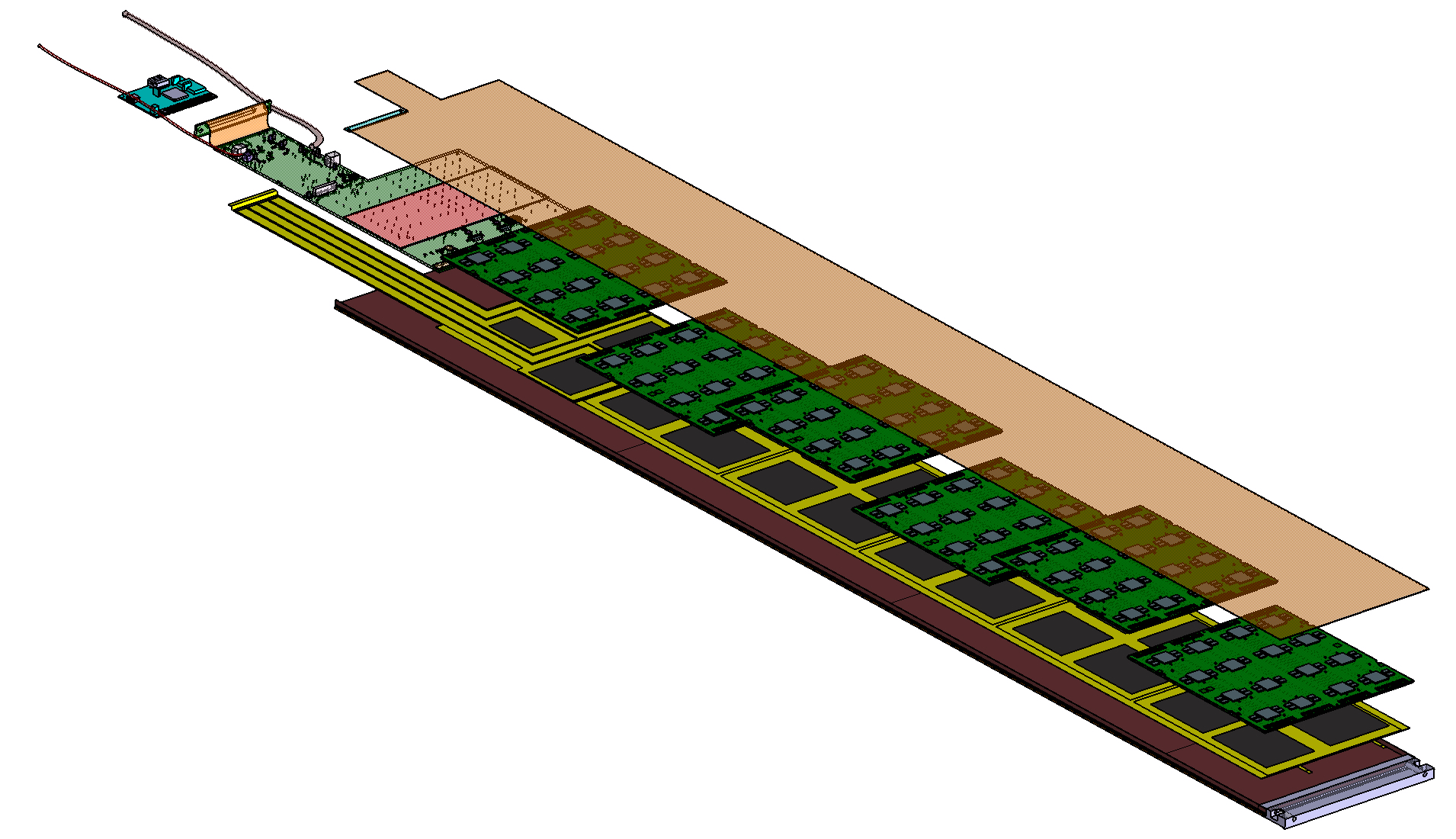}} 
\resizebox{0.45\textwidth}{!}{\includegraphics{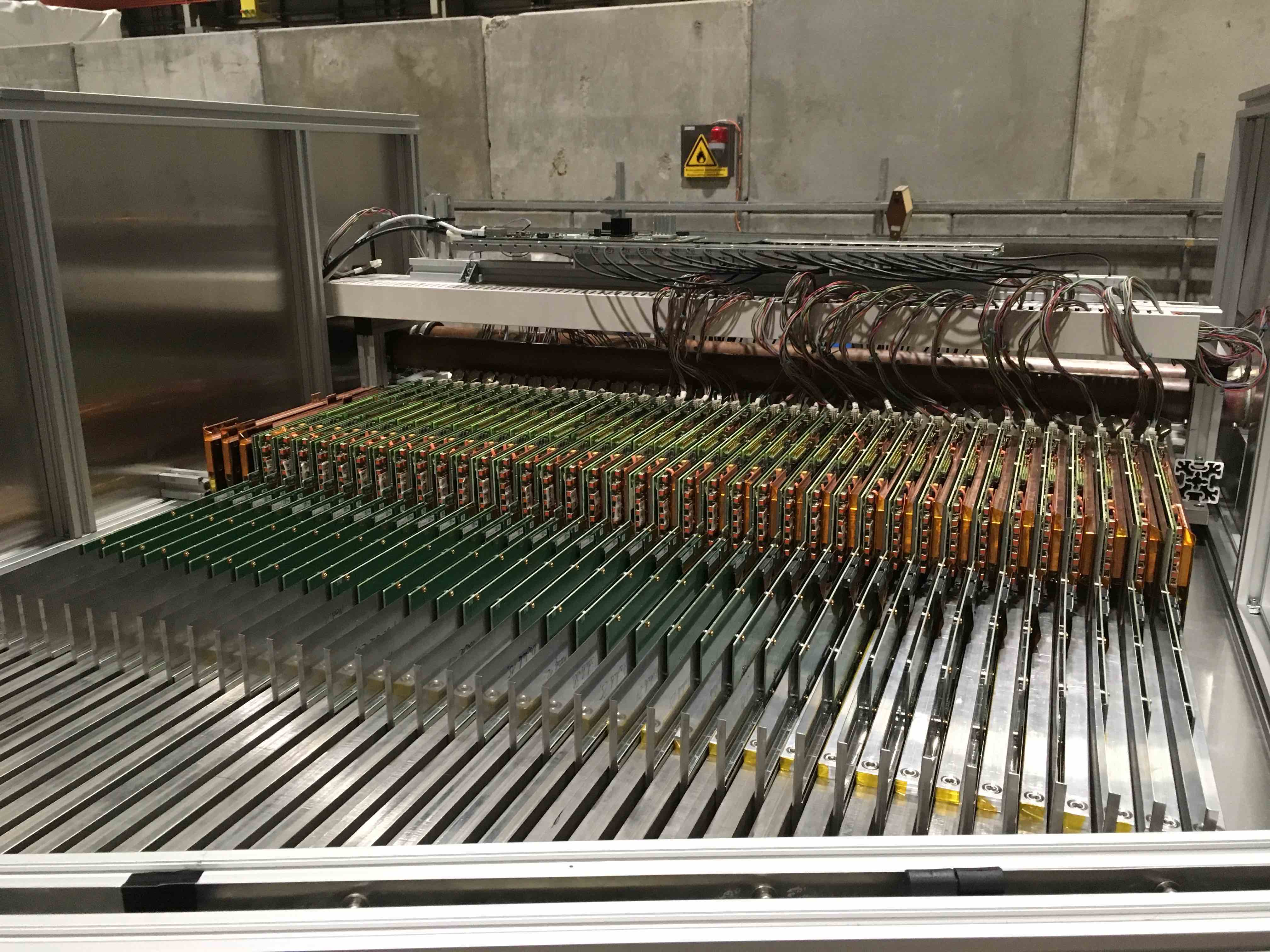}}
    \caption{Left: Active layer stack-up of the silicon-tungsten ECAL in the ILD design. The Figure shows the supporting structure including a tungsten absorber element, sensors, embedded front-end electronics as well as an external interface card \cite{Brient:2018ydi}.
%    Reproduced with permission from J.-C.\ Brient,  dd.mm.2021.
    Right: Highly granular scintillator-tile / steel hadronic calorimeter technological prototype of the CALICE collaboration, showing %one active layer with the scintillator tiles mounted on circuit boards housing the very front-end electronics and the photon sensors (left), and 
    the absorber structure with the readout interfaces for the active elements.
    }
    \label{fig:hg-ecal-ahcal}
\end{figure}

For the second generation, ultra-compact-integration solutions have been developed, which minimise the active gap widths, thus the effective Moli\`ere radius characterising the  transverse electromagnetic shower extension in the ECAL. These compact solutions - assuming a fixed hadronic interaction depth of the HCAL - also lead to a smaller calorimeter outer radius, which drives the cost of solenoid coil and return yoke in the barrel section of the detector.
The CALICE solution, adopted by ILD~\cite{ILD:2020qve} and depicted in Fig.~\ref{fig:hg-ecal-ahcal}, foresees carbon-fibre based alveolar structures with pairs of active elements attached back-to-back to an I-beam shaped tungsten plate of 1.9~mm thickness.
The silicon sensors with square pads are connected to the readout PCBs with conductive glue. 
SiD has developed a more aggressive integration concept~\cite{Barkeloo:2019zow} with ASICs directly bonded to the silicon sensors and connected to hexagonal pads via traces on the silicon.
Prototypes with up to $\sim 10$ layers have been tested for both concepts.

The CALICE SiPM-on-tile HCAL is a self-supporting stainless steel structure with 19~mm thick absorber plates and minimal un-instrumented zones, interleaved with active elements inserted as cassettes, which add another mm to the absorber thickness. The cassettes contain readout units, PCBs that hold the SiPMs, readout ASICs and LEDs for calibration, and onto which injection-moulded tiles wrapped in ESR foil are glued. 
%see Fig.~\ref{fig:hg-ecal-hcal}.
The width of the active gap is 6.5~mm, including the 3~mm thick scintillator. 
A prototype~\cite{Sefkow:2018rhp} (Fig.~\ref{fig:hg-ecal-ahcal}) with 38~layers and 21888~SiPMs has been tested in 2018 at the CERN SPS. 

The highly granular silicon-tungsten and scintillator-steel technologies are currently being applied in the endcap calorimeter upgrade of CMS~\cite{collaboration:2017gbu} for the high-luminosity phase of the LHC. 
This brings additional challenges in terms of radiation tolerance, cooling and data rates.
The channel counts and instrumented areas are at an intermediate level with respect to a full $e^+e^-$ collider detector - e.g.\ 600~m$^2$ of silicon sensors, and 240000 SiPMs - and will represent an important step in scaling production, quality control and calibration techniques, upon which an FCC-ee detector could build. 
However, the integration solutions cannot be transferred. 
The high readout bandwidth and the spatial constraints of an existing detector require to integrate into the active gaps, not only the front-end electronics, but also the first level data concentrators and the power converters, which leads to a less compact structure than that conceived for a barrel calorimeter at FCC-ee. 

\subsection{Conceptual Implementation: CLD}
While the CALICE prototypes have mainly followed the ILD detector concept~\cite{ILD:2020qve} at the ILC, the technologies have also been  adopted for the CLIC detector~\cite{AlipourTehrani:2017gov}. 
Based on the CLICdet design, the CLIC-like detector model, CLD~\cite{Bacchetta:2019fmz}, has been adapted to match the experimental conditions and physics requirements at FCC-ee.
The CLD ECAL has 40 identical silicon-tungsten layers and a total depth of 23~$X_0$. 
The silicon area of about 4000~m$^2$ is segmented into 160M cells.
The HCAL has 44 scintillator-steel layers corresponding to 5.5~$\lambda_I$.
The tiles cover an area of about 8000~m$^2$ and are read out by 9M SiPMs. 

\begin{figure}
    \centering
%    \sidecaption
%\includegraphics[width=0.5\textwidth]{figures/slab_test.jpg}  
\resizebox{0.5\textwidth}{!}{\includegraphics{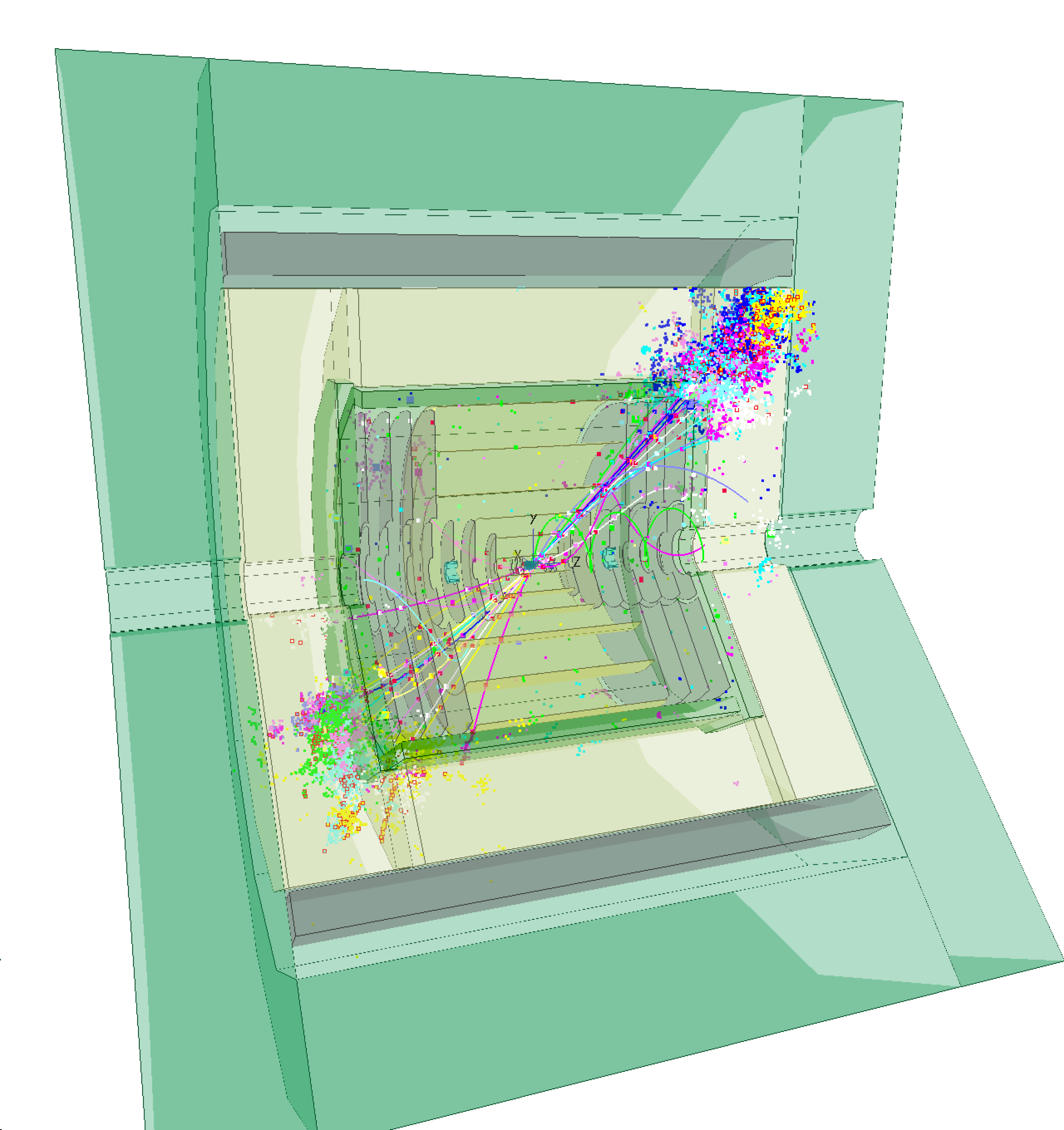}} 
    \caption{Event display in the CLD detector for a 
    $\mathrm{Z}/\gamma^*\rightarrow q\bar{q}$
    %\PZgstarToqq 
    event, with $m_{\PZ}$ = 365 GeV.
    From~\cite{Bacchetta:2019fmz}.
    }
    \label{fig:CLDjets}
\end{figure}

The energy resolution for single photons has a stochastic term of $15\%/\sqrt{E({\rm GeV})}$ in the range 5 - 100~GeV. The particle-flow jet energy resolution is 4.5\% at 50 GeV and below 4\% for energies of 100~GeV and higher. 
This results in a W-Z separation power of $2.5\,\sigma$ for boson energies of 125~GeV.
A simulated di-jet event is shown in Fig.~\ref{fig:CLDjets}. 

\subsection{Challenges and Future Developments}
In view of the challenges imposed by the tremendous statistical power of the FCC-ee on the control of systematic effects, in particular when running at the Z pole, it is mandatory to continue driving the refinement of shower simulation models and their validation using highly granular prototype data. 
The CALICE collaboration plans to continue their test beam program. 
With a fully commissioned HCAL readout, an intrinsic time resolution below 1~ns becomes possible. This will enable unprecedented studies of the shower evolution in space and time and to  explore the use of timing information for the reconstruction of topology and energy. 
A tungsten absorber structure is available to be used instead of steel and to validate simulations of hadron interactions in the preferred ECAL absorber material with fully contained showers. 
Once an ECAL prototype is fully instrumented, the combined performance of the ECAL-HCAL system will also be studied. 

The active detector elements currently at hand, like silicon sensors and SiPMs, meet the requirements; nevertheless developments in industry need to be followed. 
For example, SiPMs with smaller pixels and larger dynamic range become possible, but their reduced gain will also require more sensitive front-end electronics. 
In view of the channel counts which are more than an order of magnitude higher than in the upgraded CMS endcap calorimeter, more work on the scalability of production techniques will be needed, and different concepts will have to be followed, e.g.\ the so-called mega-tile arrays for the scintillator.   

The main challenge will be the development of an electronics system for continuous and dead-time free readout, and to address the implications for system integration. 
New front-end ASICs for energy and time measurements are needed in a common architecture for ECAL and HCAL. 
Data concentrators need to sustain much larger throughputs than the existing systems and still be highly compact, to minimise  dead zones and their impact on systematics. 
Both front-end and concentrator electronics will need active cooling, which introduces additional requirements for space for services. 
Realistic solutions have to be designed for active layers, interfaces and cooling, and should be prototyped. 

The development of integration solutions has to go hand in hand with detailed simulation studies, for example to re-optimise the absorber structure with the inclusion of copper cooling plates à la CMS. 
The CLD concept is still undetermined in some basic questions regarding the overall architecture, the segmentation of the detector into modules and the detailed design of the barrel-endcap transition.  
Signals from the embedded front-end in the barrel can be routed either along axial paths (parallel to the beam line, like in the ILD scintillator HCAL) towards interfaces at the end of the barrel, or in a tangential direction (like in the ILD ECAL) to interfaces in the gap between ECAL and HCAL.
Studies of different detector configurations in simulations with realistic assumptions on interfaces and services, validated by prototypes, will have to provide input to such key decisions. 

Finally, new ideas may be followed, for example the addition of spectral or ps timing information for the application of dual-readout methods, thus combining several of the approaches presented here.

\section{Noble-Liquid Calorimetry}
\label{sec:noble-liquid-calo}
%As outlined in Sec.~\ref{sec:intro}, the FCC-ee has an ambitious physics program at various center-of-mass energies~\cite{Mangano:2651294} which can be translated into stringent detector requirements for the FCC-ee experiments. The role of calorimetry will be to complement the tracking system in an optimal particle-flow event reconstruction. Furthermore, several B-physics measurements with $\pi^0$'s in the final states (e.g. $\mathrm{B_s}\rightarrow\mathrm{D_sK}$ in modes with $\pi^0$'s) will profit from excellent photon resolution of $\sigma_E/E\approx 5\,\%/\sqrt{E}$ and measurements with $\tau$ decays will benefit from an energy measurement of photons down to well below $1\,\mathrm{GeV}$. On top of that the requirements of particle identification will call for high transverse granularity. 

Noble-liquid calorimetry was successfully used in many high-energy experiments (e.g. E706 at FNAL, R806 at ISR, D0~\cite{D0:1993bqa}, H1~\cite{ANDRIEU1993460}, NA48~\cite{Unal:2000eu}, ATLAS~\cite{CERN-LHCC-96-041}, SLD~\cite{Benvenuti:1989ht}) due to its excellent energy resolution, linearity, stability, uniformity and radiation hardness. While radiation hardness is not a concern for lepton colliders, all other properties are clearly essential for high precision measurements, e.g. at the Z-pole, but also for the planned Higgs measurement program. 
As an example, at the Z pole, typically $10^{11}$ $\mathrm{Z}\rightarrow\mu^+\mu^-$ or Z$\rightarrow\tau^+\tau^-$ decays and $2\times 10^{12}$ hadronic Z decays will enable measurements with a statistical uncertainty up to 300 times smaller than at LEP, from a few per mil to $10^{-5}$. Such unprecedented statistical precision will have to be complemented by an extremely well controlled systematic error which requires an excellent understanding of the detector and the event reconstruction.  
A highly uniform, linear and stable measurement in the calorimeters will be a prerequisit to achieve this ambitious goal.

\subsection{Layout Optimisation for FCC-ee}
Recently, highly granular noble-liquid sampling calorimetry was proposed for a possible FCC-hh experiment~(\cite{Benedikt:2651300},~\cite{aleksa2019calorimeters} and~\cite{Aleksa:2020qdy}). It has been shown that - on top of its intrinsic excellent electromagnetic energy resolution - noble-liquid calorimetry can be optimised in terms of granularity to allow for 4D imaging, machine learning or - in combination with the tracker measurements - particle-flow reconstruction.  

Studies have started to adapt noble-liquid sampling calorimetry for an electromagnetic calorimeter of an FCC-ee experiment. Such an electromagnetic calorimeter could then be complemented by a CLD-style hadron calorimeter made of steel absorber plates, interleaved with scintillating tiles read out by SiPMs. The solenoid coil could either be located outside the hadron calorimeter, at a radius $r$ of $\approx 3.9\,\mathrm{m}$, or - provided the coil can be made thin enough - inside the noble-liquid calorimeter, at a radius $r$ of $\approx 2.1\,\mathrm{m}$, and possibly housed inside the same cryostat as the electromagnetic calorimeter. R\&D on thin carbon-fibre cryostats and thin solenoid coils as well as R\&D on high density signal feedthroughs has started in the framework of the CERN EP R\&D program~\cite{Aleksa:2649646}.

\begin{figure}
\centering
\resizebox{0.5\textwidth}{!}{\includegraphics{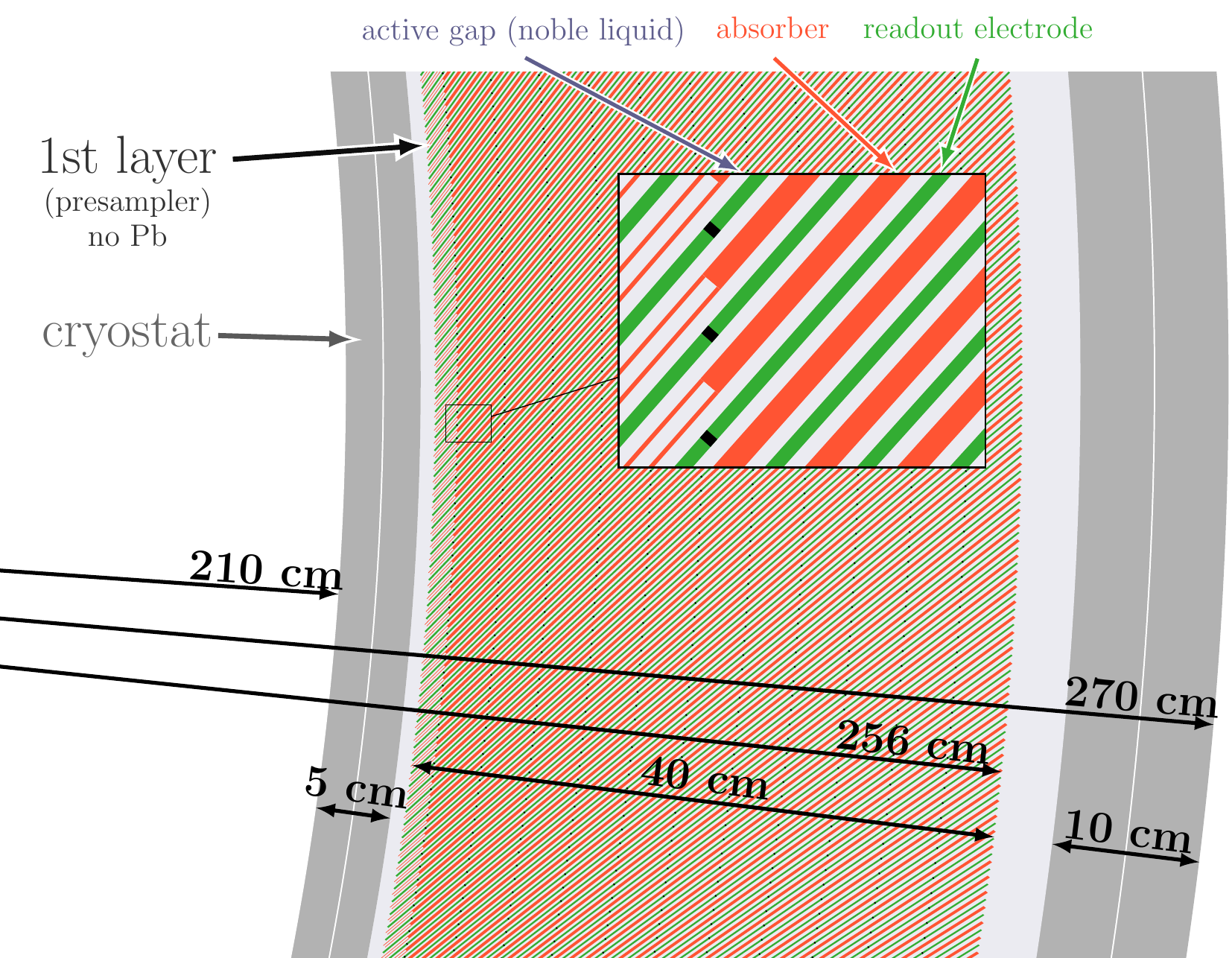}}
\caption{A noble-liquid sampling calorimeter for an FCC-ee experiment.}\label{fig:calo-fccee}
\end{figure}

Figure~\ref{fig:calo-fccee} shows a possible noble-liquid calorimeter adapted to the central region of an FCC-ee experiment, a cylindrical stack of absorbers, readout electrodes and active gaps with an inner radius of 2.1\,m, compatible with an IDEA-style tracking system (silicon vertex detector and drift chamber, see~\cite{Benedikt:2651299}). Such a configuration using liquid argon (LAr) as active material, with 1536 lead/steel absorbers of 2\,mm total thickness ($100\,\mu\mathrm{m}$ steel sheets glued onto each side of the lead absorbers), 1.2\,mm thick readout electrodes and a total depth of 40\,cm, will lead to an effective total thickness of $\sim 22$~radiation lengths, $X_0$, and a Moli\`ere radius of $R_M\approx 4\,\mathrm{cm}$. Tungsten absorbers or liquid krypton (LKr) as active material are interesting options due to the resulting smaller radiation length and smaller Moli\`ere radius which will lead to smaller showers and hence better separation of close-by particles with potentially positive impact on particle identification and particle-flow reconstruction. Studies have started to identify the best solution for an FCC-ee calorimeter. The stack of absorbers, active gaps and readout electrodes will be housed in a cryostat to reach cryogenic working temperatures. 
The electrodes, as well as the absorber plates, are arranged radially but azimuthally inclined by $\sim 50^{\circ}$ with respect to the radial direction, as shown in Fig.\,\ref{fig:calo-fccee}. This ensures that services on the inner and outer radius can pick-up the signals without creating any gaps in acceptance and allows for a high sampling frequency. The inclination of the plates has been chosen to ensure a uniform response in $\varphi$ for the full energy range despite the bending of particle tracks in the 2\,T magnetic field. 
Together with spacers defining the exact width of the active gaps, high mechanical precision and hence minimal impact on the energy resolution and uniformity can be achieved with this relatively simple structure. 
The granularity of each longitudinal compartment can be optimised according to the needs of particle-flow reconstruction and particle ID. Currently, a granularity of $\Delta\theta\times\Delta\varphi = 2.5\,\mathrm{mrad}\times 8.2\,\mathrm{mrad}$ ($5.4\,\mathrm{mm}\times 17.7\,\mathrm{mm}$) is foreseen in the first calorimeter compartment to optimise the $\pi^0$ rejection. 

In order to achieve high lateral granularity (in $\theta$), the signals will need to be brought to the edges of the electrodes via strip lines, the readout electrodes will therefore need to consist of several layers. A solution has been worked out using seven-layer printed circuit boards (PCBs), with the following layer attributions (from outside in): 
\begin{itemize}
\item The two outermost layers will be HV layers connected to HV power supplies outside the cryostat via high resistances. Together with the grounded steel surfaces of the absorbers, they will provide the electric drift field of $E_\mathrm{drift}\approx 1\,\mathrm{kV/mm}$ in the active gaps. 
\item Two layers of signal pads, $\sim 100\,\mu\mathrm{m}$ below the HV layers, will pick up the signals of the charges drifting in the drift gaps. The necessary granularity in $\theta$ will define the sizes of these pads. 
\item Two layers of ground shields will shield the signal pads from the central signal traces to avoid cross talk. 
\item The signal traces in the central layer (connected with vias to the signal pads) will bring the signals to the outer edges of the electrodes. These signal traces will form transmission lines (strip lines) together with the ground shields with an impedance matched to the input impedance of the preamplifiers. 
\end{itemize}
Such multi-layer readout electrodes will lead to increased cell capacitances potentially leading to higher series noise. For the electrodes proposed here, 2 to 10\,MeV per readout cell (at the EM scale) were estimated depending on many parameters, such as the cell size, the exact electrode design, and the electronics time constants. These low noise values obtained assuming preamplification outside the cryostats can be further improved by using cold preamplifiers inside the noble-liquid bath. Both options are studied at the moment. It is foreseen that one calorimeter cell would extend in $\phi$ over one-to-four electrodes, depending on the longitudinal compartment, which will be achieved by summing the signals on the edges of the electrodes. The granularity per longitudinal compartment will be defined by physics requirements such as $\pi^0$ identification and particle-flow reconstruction. It should be noted that the expected MIP energy deposit per double gap is $\sim 1.4\,\mathrm{MeV}$ which - correcting for a sampling fraction of $\sim 1/6$ - corresponds to a cell deposited energy of $\sim 8\,\mathrm{MeV}$. Similar to Si-based calorimeters, it will therefore be possible to track single particles in the calorimeter even before the shower starts. 

In the design described above, the active gaps between two absorbers are radially increasing from $2\times1.2\,\mathrm{mm}$, at the inner radius, to $2\times2.4\,\mathrm{mm}$, at the outer radius, leading to a sampling fraction changing with depth. Due to the longitudinally segmented readout, the shower profile will be measured for each particle, and an energy calibration, based on simulations and complemented by an $E/p$ cross calibration with the tracker, will correct for the radially dependent sampling fraction. 
%It has been shown~\cite{aleksa2019calorimeters}, that with a minimum of eight longitudinal compartments, the effect of the changing sampling fraction on the energy resolution is negligible. For FCC-ee at up to 12 longitudinal compartments are envisaged to ease particle flow reconstruction and particle identification. 
The first longitudinal compartment is realised without any absorber and will serve as a pre-sampler to be able to correct for energy lost upstream. This is especially important for low-energetic particles which loose a large fraction of their energy in the dead material in front of the calorimeter. It has been shown in~\cite{aleksa2019calorimeters} that such a correction is important for a linear energy response and improves the resolution for particles below 20\,GeV by more than 30\,\%.

The design described above will be optimised in the coming months and years to adapt it to the performance requirements by FCC-ee. Work on the readout PCBs has also started with the goal to minimise noise while achieving the necessary granularity for particle-flow reconstruction and particle ID (e.g. $\pi^0$ rejection). It is further planned to produce such readout electrodes to validate the concept in a small test calorimeter. This work as well as R\&D on high-density signal feedthroughs, thin carbon-fibre cryostats and thin solenoid coils is part of the CERN EP R\&D program~\cite{Aleksa:2649646}. 

\subsection{Performance}
The performance of such a calorimeter has been evaluated for an FCC-hh experiment~\cite{aleksa2019calorimeters,Aleksa:2020qdy}. The below quoted photon, electron and pion stand-alone performance is therefore meant to demonstrate that such a calorimeter - if further optimised - has big potential to achieve all above listed performance requirements. It should be noted that eventually particle-flow event reconstruction will be used. 
%It consists in a combination of the tracker measurement of all charged particles and the calorimetry measurement, relying on high calorimeter granularity and excellent position resolution.
At this moment, particle-flow event reconstruction is being implemented into the FCC software, but is unfortunately not yet available for performance studies.

Single-particle simulations of electrons and photons have resulted in a stochastic term of 8.2\,\%~\cite{aleksa2019calorimeters,Aleksa:2020qdy} for the standalone electromagnetic energy resolution. This value can be further improved by increasing the sampling fraction or the sampling frequency. It has been shown that the noise contribution, which dominates energy resolution at low particle energies, can be kept below 50\,MeV by optimising the cluster size. For charged particles, the reconstruction will rely on a particle-flow combination of the tracker and the calorimetry measurement, relying on an excellent position resolution. An ECAL position resolution of $<500\,\mu\mathrm{m}$ was simulated for energy deposits $>30$\,GeV. Fine lateral segmentation is also essential for the required $\pi^0$ rejection\footnote{The $\pi^0$ rejection factor $R_{\pi^0}$ is defined as the fraction of the total number of $\pi^0$ divided by the number of non-rejected $\pi^0$.}, which was shown to be $R_{\pi^0}>5$ for transverse momenta up to $30$\,GeV in a deep-neural-network based analysis (assuming 90\,\% signal efficiency)~\cite{aleksa2019calorimeters}.

The single-$\pi^-$ resolution has been obtained from simulation using a simple hadron calibration, the so-called benchmark method~\cite{aleksa2019calorimeters}. It consists in adding the simulated energy deposits in a window of defined size, taking into account the different hadronic response of ECAL and HCAL, and correcting the obtained energy for the energy lost in the dead material in between the calorimeters. In the FCC-hh simulation, a stochastic term of 48\,\% (44\,\%) for single $\pi^-$ was achieved in this way for $B=4$\,T ($B=0$\,T), respectively. Exploiting the full 4D imaging information, it was demonstrated that a deep-neural-network analysis can achieve a single-$\pi^-$ resolution stochastic term of 37\,\% ($B=4$\,T). It should be noted that this remarkable result relies on the calorimeter measurement alone, the planned particle-flow reconstruction combined with the tracker will further substantially improve the hadronic resolution.

\section{Crystal Calorimetry}
\label{sec:crystals}
Crystal calorimeters have a long history of pushing the frontier on high-resolution electromagnetic
(EM) calorimetry for photons and electrons.
More recently, with the advent of the SiPM photodetection technology, segmented crystal calorimeters incorporate and achieve new performance benchmarks for precision timing, particle identification and $e/h$ response compensation through dual readout.  These extended capabilities are central to the FCC-ee physics program where a high level of precision is required to comprehensively measure and identify all particles forming the event from a wide range of processes, from rare Higgs decays, heavy flavour and $\tau$-lepton decay chains, to low systematic electroweak measurements.  Segmented crystal calorimeters are only one part of a complete measurement system but studies, in concert with a low-mass tracking system and a dual-readout fibre hadron calorimeter, highlight which parameters are most important for the combined detector performance~\cite{Lucchini_2020}.

Inorganic crystals have intrinsically low response to neutrons and low-mass nuclear fragments due to inefficient momentum transfer to the heavy materials that form the crystal lattice.  However, the bias towards low response is well measured by the high EM response in the form of both scintillation and \v{C}erenkov light, providing an accurate dual-readout compensation.
Considering first the neutral components of the $ZH$ events, those not captured by the tracking system, adding dual readout to the rear compartment of a segmented crystal provides an excellent neutral hadron resolution of 30\%/$\sqrt{E} \oplus 2$\% when combined with the dual-readout fibre calorimetry, while achieving 3\%/$\sqrt{E}$ for low energy photons~\cite{Lucchini_2020}.  This combination balances the need to bring down the leading contribution to the overall jet energy resolution, originating primarily from $K_L$'s with an average energy of approximately 5.5~GeV, while at the same time maximising the EM resolution.
The EM resolution for low energy photons is a sub-dominant component of the total jet energy resolution but the ability to resolve and correctly pair photons into $\pi^0$'s, the underlying hadron momenta, resolves an important particle assignment ambiguity when forming jets in 4- and 6-jet $ZH$ events.  Similarly, the effects of dead material from the finite inner radius of the solenoid can be mitigated by placing the solenoid between the crystal calorimeter and the dual-readout fiber calorimeter, where low energy EM particles are directly measured with crystals with a minimum of dead material losses.

\begin{figure}[ht]
\centering
\resizebox{0.5\textwidth}{!}{\includegraphics{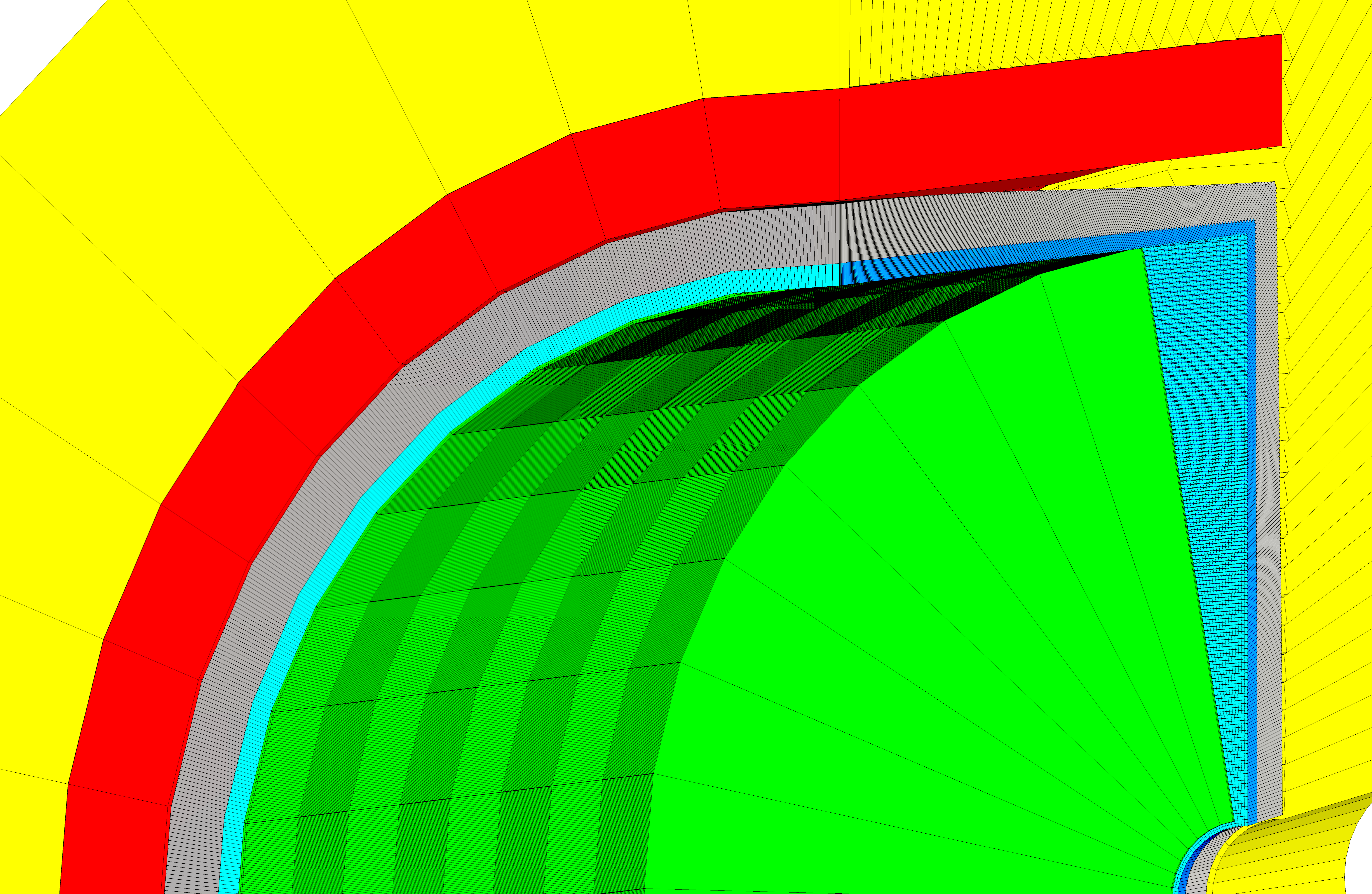}}
\resizebox{0.27\textwidth}{!}{\includegraphics{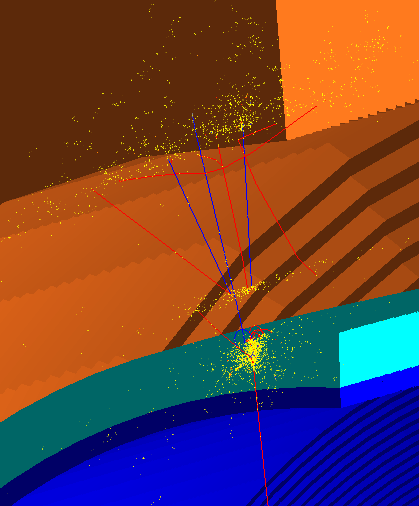}}
\caption{A segmented crystal calorimeter integrated into an FCC-ee experiment.  ({\it left}) The precision timing layers (green) are followed by projective crystals longitudinally segmented into front (blue) and rear (grey) compartments.  The rear crystals are instrumented with dual readout and are surrounded by a solenoid (red) in the barrel region and hermetically by a dual-readout fibre calorimeter (yellow).  ({\it right}) A 10~GeV pion shower through the crystal calorimeter option of the IDEA detector~\cite{ALY2020162088}.}\label{fig:calo-crystals}
\end{figure}

The inner tracking system measures the charged hadron momenta with a higher resolution than can be achieved with calorimetry but, to fully benefit from this increased resolution, particle-flow algorithms rely on particle identification and low-ambiguity cluster assignment from the calorimeter.  A segmented crystal calorimeter with precision timing layers can tag arrival times for MIPs to better than 20~ps.  Precision timing fills the well-known gap in $dE/dx$ particle identification from the minimum region of the Bethe-Bloch energy-loss function.  Charged hadrons are further characterised by the penetration depth before showering begins, determined from the delayed energy profiles in longitudinally segmented crystal readout, and the relative amount of \v{C}erenkov (C) to scintillation (S) light produced in the shower.  These ratios have been shown to be powerful tools to resolve particle identification within a segmented crystal calorimeter, acting as a linchpin between tracks and dual-readout fibre calorimeter clusters.  With crystal measurements alone, when using a front/rear ratio, transverse profile and C/S, a rejection of 99.4\% on charged pions, for a 99\% efficiency to select 10~GeV electrons, has been estimated~\cite{Lucchini_2020}.  An example layout of a segmented crystal calorimeter integrated into an FCC-ee experiment is shown in Fig.~\ref{fig:calo-crystals}.

The material budget of the inner tracker has a profound impact on the overall performance of crystal calorimeters.  Studies on the material thickness in radiation lengths show that the electrons from $Z\rightarrow e^+e^-$ decays radiate a substantial fraction of their momenta into bremsstrahlung photons.  The $Z\rightarrow e^+e^-$ mass recoil is a crucial quantity for $HZ$ associated Higgs production studies.  Compared to $Z\rightarrow \mu^+\mu^-$ decays measured with a tracking momentum resolution of 0.3\%, the resolution on the $Z\rightarrow e^+e^-$ mass recoil is approximately 3 times worse for an EM resolution of 15\%/$\sqrt{E}$ for a tracker thickness of 0.4~$X_0$.
With segmented crystal calorimetry, the recoil mass resolution of $Z\rightarrow e^+e^-$ is within 25\% of the muons for tracker thicknesses up to 0.4~$X_0$~\cite{Lucchini_2020}. For a substantially thicker tracker, the bremsstrahlung recovery does not show significant improvement on the electron momentum.

Crystal R\&D continues to be a major component of new advances in crystal calorimetry.  Bright, dense crystals, such as LYSO, and with ultra-fast rise time, with doped-BaF$_2$, are being produced in large-scale for high-precision timing detectors~\cite{Kratochwil_2020,OMELKOV2018260}.  Cutting and growth methods are providing new possibilities for segmentation at low cost.  Fast photon production processes in crystals, like \v{C}erenkov photons and intra-band luminescence, are promising avenues of exploration for timing application.  New techniques, such as 3D printing, laser growth and nano-engineered materials, are creating new types of crystal scintillator structures~\cite{TURTOS2019116613}.  The customization of SiPM parameters to match crystal parameters is an area of rapid development with many new directions to increase photodetection efficiencies, wavelength coverage, timing uniformity, dynamic range, fast recovery time and low noise performance.

Beyond these important benchmarks, the full power of segmented crystal geometries is still unknown.  A so-called ``5D Crystal'' geometry leverages multiple stacks of segmented crystal timing layers to form the entire crystal calorimeter, following the single-layer geometry of the CMS barrel timing layer~\cite{CERN-LHCC-2017-027}.  The individual positions of shower energy deposits are resolved using pattern recognition and timing.  The resulting spatial shower information is comparable to a 16-layer silicon-tungsten sampling calorimeter, without the loss in sampling fraction from the interleaved absorber material.  A key component of future crystal calorimeter developments is an optimisation procedure for particle-flow algorithms that attempts to use coarse or even non-uniform longitudinal segmentation while still retaining more information per cell due to the high, near unity, sampling fraction of homogeneous crystals.  With the unprecedented statistical power, diversity and breadth of the FCC-ee physics program, the challenges in crystal calorimeter design are even tougher, with the need to evaluate detector parameters and their potential impact on the ultimate limits on never-before-achieved measurement precisions and new physics search sensitivities.

\section{Dual-Readout Fibre Calorimetry}
\label{sec:dual-readout}
\begin{figure}
    \centering
    %\resizebox{0.5\textwidth}{!}{\includegraphics{figures/WZH-jj.pdf}} 
    \resizebox{0.7\textwidth}{!}
        {\includegraphics{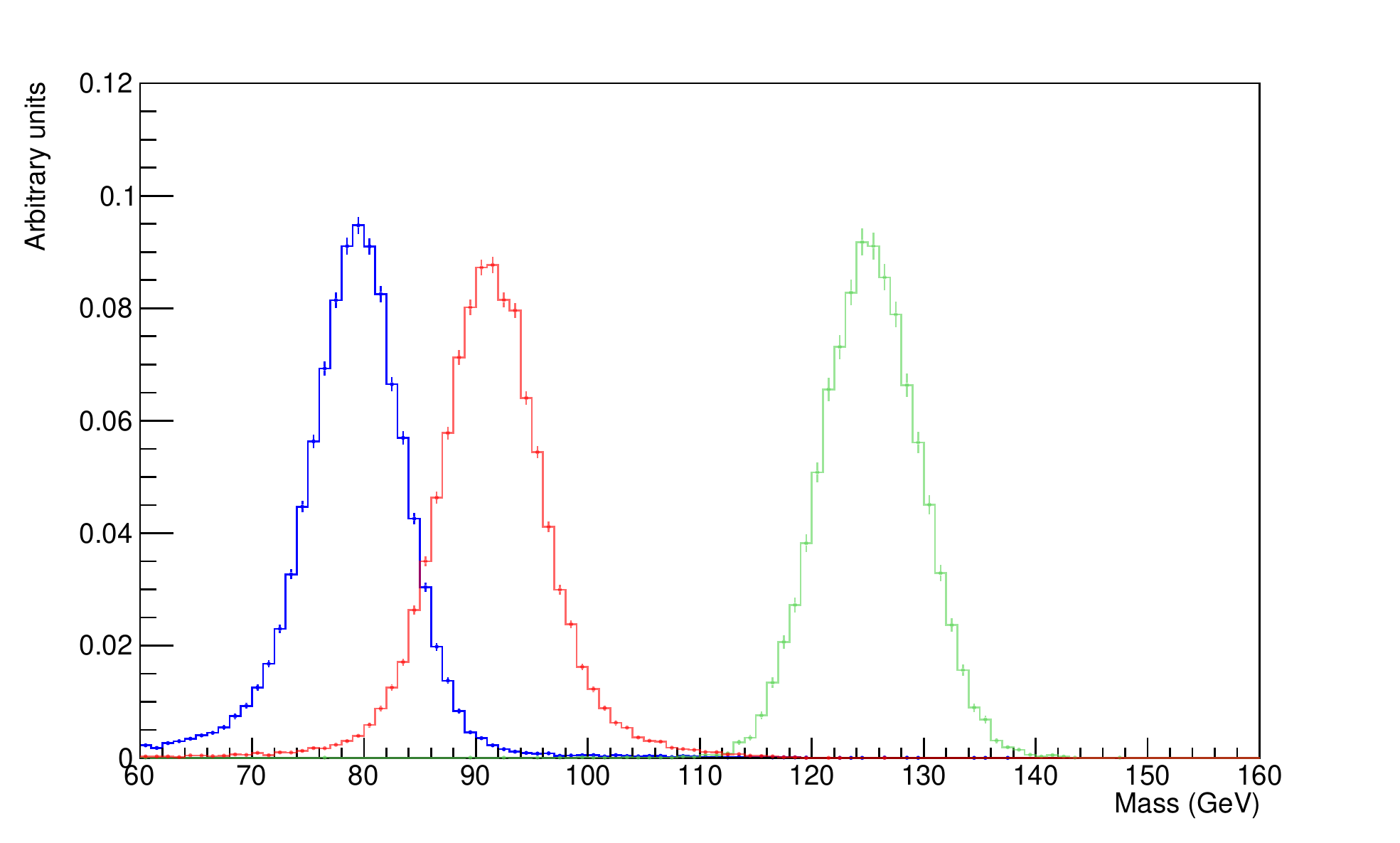}} 
    \caption{Reconstructed invariant mass distributions for  
    $W/Z/H \rightarrow jj$ events in a dual-readout fibre calorimeter.
    }
    \label{fig:WZH-jj}
\end{figure}

To optimally benefit from the large datasets that will be available at FCC-ee and fully exploit its physics program, the intrinsic measurement resolutions and event-to-event information have to be substantially increased. The 20-year-long $R\&D$ program on Dual-Readout Calorimetry (DR, DRC) of the DREAM/RD52 collaboration~\cite{DRC_Wigmans,RD52_emshow,RD52_emperf,RD52_hadperf,RD52_MC,RD52_PID,RD52_SiPM,IDEA_SiPM} shows that, with a detector fully calibrated at the EM scale, the independent readout of scintillation (S) and \v{C}erenkov (C) light allows the cancellation of the effects of the fluctuations in the EM fraction of hadronic showers. The DR fibre-sampling approach brings the stochastic term down close to or even below $30\%/\sqrt{E}$ through a high sampling frequency and the integration of the shower over its longitudinal development. The latter, in addition, leads to a less-noisy information.
With an ideal detector (with an energy resolution of $\sim 30\%/\sqrt{E}$), the expected separation reachable for the $W/Z/H \rightarrow jj$ peaks is shown in Fig.~\ref{fig:WZH-jj}.
Stand-alone results show as well excellent particle-ID performance and competitive EM energy resolution.

The advancements in solid-state light sensors such as SiPMs have opened the way for highly granular fibre-sampling detectors with the capability to resolve the shower angular position at the mrad level or even better.
In the present design, 1-mm diameter fibres are placed, at a distance (apex to apex) of 1.5-2 mm, in a brass absorber matrix (copper, iron and lead being the alternative materials under consideration). This means that the lateral segmentation could be pushed down to the mm level, largely enhancing the resolving power for close-by showers, with a significant impact, for example, on channels like $\tau \rightarrow \rho \nu$.

The high PDE of SiPMs should permit to obtain light yields of O(100) p.e./GeV for both the S and C signals, which can guarantee an EM resolution close to $10\%/\sqrt{E}$. In the above geometry, the lower limit, as set by the sampling fluctuations, is around $(8-9)\%/\sqrt{E}$.
Readout ASICs providing time information with $\sim$~100 ps resolution may allow the reconstruction of the shower position with $\sim$~5 cm of longitudinal resolution.

On the other hand, the large number and density of channels call for an innovative readout architecture for efficient information extraction. Both charge-integrator and waveform-sampling ASICs are available on the market and candidates for the first tests have been identified (the Weeroc Citiroc 1A charge integrator and Nalu Scientific system-on-chip digitisers). At the time of writing, a first implementation of a scalable readout system is almost ready for testing in a time scale of few months.
Looking further ahead, digital SiPMs (dSiPMs) should allow significant simplification of the readout architecture but the technology does not yet appear mature enough. A specific $R\&D$ program has been submitted for approval.

The mechanical assembly and integration of a system with $O(10^8)$ sensitive elements require the development of a robust and engineered procedure.
A scalable mechanical solution, that should work for both non-projective and projective modules, has been defined. Based on the gluing of capillary tubes, it is being exploited for building a small ($\sim 10 \times 10 \times 100\ cm^3$) EM prototype to be tested with beam (yet within few months). The preliminary results are very positive and this approach is presently the basis of a project for the construction of a hadronic prototype (of size $\sim 60 \times 60 \times 200\ cm^3$) in 3-4 years, depending on funding approval. Alternative approaches, in particular with 3D printing, are being investigated within a South Korean $R\&D$ project.

The performance, in the reconstruction of the properties of both hadronic and EM showers, is good enough to open the possibility to exploit a single integrated dual-readout fibre-sampling solution, for the calorimetric system of an FCC-ee experiment. This is the baseline choice in the IDEA~\cite{IDEA_tb1} detector concept. 

The huge amount of information made available by the fibre SiPM readout should be likely take advantage of deep-learning algorithms, in order to be maximally exploited. The preliminary performance in the identification of $\tau$-decay final states, using calorimetric information only, looks very promising. With reduced (full) fibre information the average classification accuracy was estimated to be $\sim 90\%$ ($> 99\%$).

\section{Conclusions}
\label{section:conclusion}
Calorimetry will play a crucial role for the particle reconstruction, identification and measurement in any FCC-ee experiment. The next-generation collider experiments will be optimised to combine measurements across sub-detectors to achieve unprecedented accuracy in identifying and measuring with high resolution all particles in the event.
Calorimeters will be tasked with
%to provide high resolution energy, angle and resolving power not only for neutral particles, but to further aid in the unambiguous separation of charged particles by strengthening the matching criteria through timing, granularity, shower identification and higher resolution.  The two primary classes of calorimeters in current experiments have separate optimisations for 
%particle flow event reconstruction, based on the principle of measuring each particle individually and by using the best measurement for each, and for high resolution energy measurement using dual read-out calorimetry, which achieves excellent jet resolution by correcting for the event-to-event fluctuations of the electromagnetic shower fraction. The increased demands of the FCC-ee physics program are pushing calorimeter designs to further incorporate multiple complementary techniques to 
maximising information to augment the event description in step with developments across all sub-detectors.  We have discussed the state of the art of different technologies that are well suited to be used in an FCC-ee experiment. Table~\ref{tab:sum-resolution} summarizes the expected energy resolution for the different technologies.

\begin{table}[]
\resizebox{\textwidth}{!}{%
\begin{tabular}{l|ccccc}
\hline
Detector technology & E.m. energy res. & E.m. energy res. & ECAL \& HCAL had. & ECAL \& HCAL had. & Ultimate hadronic \\
(ECAL \& HCAL) & stochastic term & constant term & energy resolution & energy resolution & energy res. incl. PFlow \\
& & & (stoch. term for single had.) & (for 50\,GeV jets) & (for 50\,GeV jets) \\ 
\hline\hline
Highly granular & \multirow{3}{*}{15 -- 17\,\%~\protect\cite{Anduze:2008hq,Bacchetta:2019fmz}} & \multirow{3}{*}{$1$\,\%~\protect\cite{Anduze:2008hq,Bacchetta:2019fmz}} & \multirow{3}{*}{$45-50$\,\%~\protect\cite{CALICE:2009vtf,Bacchetta:2019fmz}} & \multirow{3}{*}{$\approx 6$\,\% ?} & \multirow{3}{*}{4\,\%~\protect\cite{Bacchetta:2019fmz}} \\
Si/W based ECAL \& & & & & & \\
Scintillator based HCAL & & & & & \\
\hline
Highly granular & \multirow{3}{*}{8 -- 10\,\%~\protect\cite{CERN-LHCC-96-041,aleksa2019calorimeters,Abat_2010}} & \multirow{3}{*}{$<1$\,\%~\protect\cite{CERN-LHCC-96-041,aleksa2019calorimeters,Aad_2019}} & \multirow{3}{*}{$\approx 40$\,\%~\protect\cite{aleksa2019calorimeters,Aleksa:2020qdy}} & \multirow{3}{*}{$\approx 6$\,\% ?} & \multirow{3}{*}{3 -- 4\,\% ?} \\
Noble liquid based ECAL \& & & & & & \\
Scintillator based HCAL & & & & & \\
\hline
Dual-readout & \multirow{2}{*}{11\,\%~\protect\cite{Antonello_2020}} & \multirow{2}{*}{$<1$\,\%~\protect\cite{Antonello_2020}} & \multirow{2}{*}{$\approx 30$\,\%~\protect\cite{Antonello_2020}} & \multirow{2}{*}{4 -- 5\,\%~\cite{Antonello:2020htm}} & \multirow{2}{*}{3 -- 4\,\% ?} \\
Fibre calorimeter & & & & & \\
\hline
Hybrid crystal and & \multirow{2}{*}{3\,\%~\protect\cite{Lucchini_2020}} & \multirow{2}{*}{$<1$\,\%~\protect\cite{Lucchini_2020}} & \multirow{2}{*}{$\approx 26$\,\%~\protect\cite{Lucchini_2020}} & \multirow{2}{*}{5 --  6\,\%~\cite{Lucchini_2020,lucchinieps}} & \multirow{2}{*}{3 -- 4\,\%~\protect\cite{lucchinieps}} \\
Dual-readout calorimeter & & & & & \\
\hline
\end{tabular}}
\caption{Summary table of the expected energy resolution for the different technologies. The values are measurements where available, otherwise obtained from simulation. Those values marked with "?" are estimates since neither measurement nor simulation exists.
}
\label{tab:sum-resolution}
\end{table}

% We have shown that there are several promising candidates of calorimeter technologies that are expected to allow us to fully exploit the FCC-ee physics program. 
We have shown that there are several promising candidates of calorimeter technologies proposed for the FCC-ee physics program. Their differences are the result of the original principles they are built from, however they are all going into the direction of exploiting the technological advancements for pushing the spatial granularity and the timing performance well beyond what presently achieved. Such a variety of approaches is a guarantee for an optimal mining of the FCC-ee physics potential and, moreover, there are no showstoppers preventing the integration of the different solutions in a single design. 
Intensified R\&{}D effort will be essential to further develop all the above concepts and build prototypes for each of the proposed technologies.

%
% For  figures use
%\begin{figure*}
%\vspace*{5cm}       % Give the correct figure height in cm
%\includegraphics{leer.eps}
%\end{figure*}
% or  this
%\begin{figure}
%\centering
%\resizebox{0.75\textwidth}{!}{%
%  \includegraphics{leer.eps}
%\vspace{5cm}       % Give the correct figure height in cm
%\end{figure}
%\begin{table}
%\centering
%\begin{tabular}{lll}
%\hline\noalign{\smallskip}
%\end{tabular}
% Or use
%\vspace*{5cm}  % with the correct table height
%\end{table}

\bibliographystyle{jhep}
\bibliography{references}
\end{document}